\documentclass[preprint,showpacs,preprintnumbers,groupedaddress,amsmath,amssymb,aps,prl]{revtex4}

\usepackage{graphicx}

\begin{document}

\title{Comment on "Direct Observation of Optical Precursors in a Region of Anomalous Dispersion"}
\author{Bruno Macke} 
\author{Bernard S\'{e}gard} 
\affiliation{Laboratoire PhLAM, Universit\'{e} de Lille I, 59655 Villeneuve d'Ascq, France}

\date{\today}
\maketitle

In a recent Letter \cite{ref1}, Jeong, Dawes and Gauthier (JDG) claim to have achieved the first direct measurement of "optical precursors" for a step-modulated field propagating through a anomalously dispersive medium. In fact such transients have been evidenced previously \cite{ref2}. They are not identifiable to precursors (unless one considers that any coherent transient propagating in a dilute medium at the velocity $c$ is a precursor) and they can be interpreted in very simple physical terms.

The Sommerfeld-Brillouin precursors occur in a medium of large optical thickness (a condition met in \cite{ref2} but not in the JDG experiments). Their signature is an evolution of the \emph{electric field} at time scales strongly deviating from the period of the incoming field. They are thus excited if and only if this field is turned on in a time at worst comparable to its period. This condition is not met in the experiments where, in addition, one observes the (slow)\emph{envelope} of the field (or the corresponding intensity profile) instead of the field itself. The transients reported in \cite{ref1,ref2} are thus not identifiable to precursors and, even with much goodwill, it is in particular impossible to recover in the exponential-like transients observed by JDG the richness of the precursors dynamics. As a further argument, we remark that the asymptotic theory, specially adapted to the study of the precursors, dramatically fails to explain the experimental results (predicted amplitudes up to $10^{10}$ times too large, inability to reproduce the oscillatory behavior at large optical thickness). Conversely, the slowly varying envelope approximation (SVEA) is perfectly adapted to the experiments where the switching time, although much shorter than the medium response-time, is very long compared to the optical period. 
	
Since the JDG experiments has been achieved in a medium of moderate optical thickness, we first determine the envelope $e(L,\tau)$ of the transmitted field  in the optically thin sample limit (we use the notations of JDG). Solving the Bloch-Maxwell equations at the first order in $\alpha_{0}L/2$, we directly get $e(L,\tau)\approx E_{0}\Theta(\tau)-E_{0}\Theta(\tau)\left(1-e^{-\delta\tau}\right)\alpha_{0}L/2$. The corresponding intensity profile $\left|e(L,\tau)\right|^{2}$ very satisfactorily fits that observed by JDG for $\alpha_{0}L=0.41$ (see their Fig.1). Our expression of  $e(L,\tau)$ illustrates a general property, namely that the transmitted wave is the sum of the incoming wave (as it would propagate in vacuum) with the secondary wave emitted by the polarization induced in the medium. This result can obviously be extended to arbitrary optical thickness. Using the expression of the medium impulse-response obtained by Crisp [3] (his Eq.33), we then get 

\begin{equation}	e(L,\tau)=E_{0}\Theta(\tau)\left[1-\alpha_{0}L\int^{\delta\tau}_{0}\frac{J_{1}(\sqrt{2\alpha_{0}Lu})}{\sqrt{2\alpha_{0}Lu}}e^{-u}du\right]\label{eq1}
\end{equation}
Contrary to Eqs. 4 and 5 of JDG, this form is valid for any time $\tau$. The first order result is retrieved by remarking that $J_{1}(x)/x= 1/2+O(x^{2})$ when $x<<1$. At moderate optical thickness, the effect of the function weighting $e^{-u}$ in Eq.\ref{eq1} is essentially to shorten the transient, as observed by JDG for $\alpha_{0}L=1.03$. When  $\alpha_{0}L>>1$, the transient becomes oscillating.
 
\begin{figure}[h]
 \begin{center}
   \includegraphics[width=8cm]{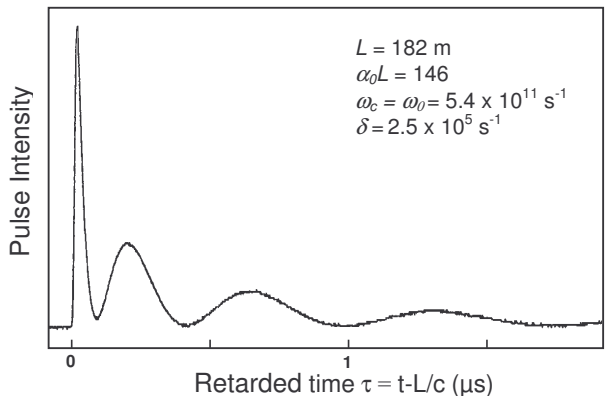}
   \caption{Direct observation of the response of an optically thick medium to a resonant step-modulated wave. This result completes that obtained by JDG at moderate optical thickness (their Fig. 1).\label{fig1}}
\end{center}
\end{figure}
Figure \ref{fig1} shows the intensity-profile  experimentally obtained for $\alpha_{0}L=146$ \cite{ref2}. It fully agrees with the prediction of Eq. \ref{eq1}. In particular the maximums exactly occur at the retarded times $j^{2}_{1,n}/(2\alpha_{0}L\delta)$ , where $j_{1,n}$ is the n$^{th}$ zero of $J_{1}(x)$. Let us remind that the oscillatory behavior of the transient is not reproduced by the current theory of precursors. Since on the contrary the transients reported in \cite{ref1,ref2} perfectly agree with the theoretical predictions by Crisp \cite{ref3}, we suggest, to avoid any confusion, to name them \emph{Crisp transients} instead of optical precursors.

\end{document}